\documentclass[pre,twocolumn,superscriptaddress,showpacs,preprintnumbers,amsmath,amssymb]{revtex4}

\newcommand{\bq}{\begin{equation}}
\newcommand{\ba}{\begin{eqnarray}}
\newcommand{\eq}{\end{equation}}
\newcommand{\ea}{\end{eqnarray}}

\newcommand {\calD} {{\cal D}}

\newcommand{\Ord}{\mathrm{O}}

\usepackage{amsfonts}
\usepackage{amssymb}
\usepackage{amsmath}
\usepackage{feynmp}
\usepackage{color}
\usepackage{graphicx}
\usepackage[pdftex,colorlin ks=true]{hyperref}

\begin{document}

%\title{Momentum space structure of  the Gray Scott Model with internal noise}
\title{Effects of intrinsic noise on a cubic autocatalytic reaction diffusion system}
\author{Fred Cooper} \email{fcooper@fas.harvard.edu}
\affiliation{Department of Earth and Planetary Sciences, Harvard University, Cambridge, MA 02138}
\affiliation{The Santa Fe Institute, 1399 Hyde Park Road, Santa Fe, NM 87501, USA}
\author{Gourab Ghoshal}   \email{gghoshal@fas.harvard.edu}
\affiliation{Department of Earth and Planetary Sciences, Harvard University, Cambridge, MA 02138}
\author{Juan P\'erez-Mercader} \email{jperezmercader@fas.harvard.edu}
\affiliation{Department of Earth and Planetary Sciences, Harvard University, Cambridge, MA 02138}
\affiliation{The Santa Fe Institute, 1399 Hyde Park Road, Santa Fe, NM 87501, USA}

\date{\today}

\begin{abstract}

Starting from our recent chemical master equation derivation of the model of an autocatalytic reaction-diffusion chemical system with reactions $U+2V  {\stackrel {\lambda_0}{\rightarrow}}~ 3 V;$ and $V {\stackrel {\mu}{\rightarrow}}~P$, $U  {\stackrel {\nu}{\rightarrow}}~ Q$,  we determine  the effects of intrinsic noise on the momentum-space  behavior of its kinetic parameters and chemical concentrations.  We demonstrate that the intrinsic noise induces $n \rightarrow n$  molecular interaction processes with $n \geq 4$, where $n$ is the number of molecules participating of type $U$ or $V$.  The momentum dependences of the reaction rates are driven by the fact that the autocatalytic reaction (inelastic scattering) is renormalized through the existence of an arbitrary number of intermediate elastic scatterings, which   can also be interpreted as  the creation and subsequent decay of a  three body composite state $\sigma =  \phi_u \phi_v^2$, where $\phi_i$ corresponds to the fields representing the densities of $U$ and $V$. Finally, we discuss the difference between representing $\sigma$ as a composite or an elementary particle (molecule)  with its own kinetic parameters. In one dimension we find that while they show markedly different behavior in the short spatio-temporal scale, high momentum (UV)  limit, they are formally equivalent in the large spatio-temporal scale,  low momentum (IR) regime.  On the other hand in two dimensions and greater, due to the effects of fluctuations, there is no way to experimentally distinguish  between a fundamental and composite $\sigma$.  Thus in this regime $\sigma$ behave as an entity unto itself suggesting that it can be effectively treated as an independent chemical species.
\end{abstract}
        
\maketitle

\section{Introduction}

Reaction diffusion (RD) systems are a versatile class of models capable of encoding a variety of  phenomena observed in Nature in areas encompassing physics, biology, ecology, chemistry and many other fields~\cite{CG09,Walgraef97,Mikhailov_1994, Grzybowski09}. Their application to chemical systems are of particular interest, as they include biologically relevant phenomena such as pattern formation and self-replication, and therefore  can be used as proxies for high-level biological systems~\cite{Pearson93,LMPS94}. While RD systems have been mostly studied from a deterministic standpoint, any faithful application to biological systems \emph{must} take into account the effects of noise, since an important facet of such systems is its exchange of matter and energy with the environment; a process which clearly brings in some amount of stochasticity.  

To reflect this, there have been recent efforts to study stochastic chemical reaction diffusion systems~\cite{LHMPM03,HLMPM03, CGPM_2013}. In particular, when such systems are coupled with \emph{external} noise it is known that there are  renormalization effects due to the fluctuations represented by the noise (by external we mean fluctuations that are not inherent to the chemistry itself). These affect for example, the strength of the chemical interactions that,  in turn, induce new interactions not originally present in the ``macro-level" chemistry \cite{HLMPM03}.  However, independently of the above, there is also some form of \emph{intrinsic} noise in the chemical system whose effects are less understood. Qualitatively, one can interpret this intrinsic noise as a manifestation of the underlying mechanisms that lead to the observed behavior of reaction diffusion systems at the level of their chemical kinetics. Because of this, it is important to understand the precise nature and effects of $intrinsic$  noise, as it might give intuition and provide hints for understanding the internal structure of the system~\cite{prlus}. 

In light of this, in this paper we seek to determine whether the inherent stochasticity in the nature of the chemical reactions themselves leads to effects similar to those induced by the external or $extrinsic$ noise. Of course this particular stochasticity is restricted by certain assumptions when we attempt to model these reactions in terms of kinetics. The most basic is that the molecules are random walkers in a $d$-dimensional space $and$ that collisions between them occur as a function of the probability of encounters between these random walkers. Most collisions are elastic and do not result in a chemical reaction, whereas comparatively few are inelastic and lead to the actual chemistry that we are interested in. The relative scarcity of the latter with respect to the former implies that the chemically interesting inelastic collisions are effectively statistically independent and therefore the \emph{chemical reactions} (occurring at large scales) are Markovian in nature.  (Note that the assumption of the chemical reactions as a Markovian process is valid only up to a resolution limit which corresponds to the \emph{mean free path} of the molecules involved in the reaction.)   

In a previous paper~\cite{CGPM_2013} we considered as a test case a generic spatially extended set of  macroscopic chemical reactions, 
\bq  
U+2V  {\stackrel {\lambda_0}{\rightarrow}}~ 3 V, ~~
V {\stackrel {\mu}{\rightarrow}}~P, ~~
U  {\stackrel {\nu}{\rightarrow}}~ Q,~~ {\stackrel {f}{\rightarrow}}~U.   
\label{reactions}
\eq
There is a cubic autocatalytic step for $V$ at rate $\lambda_0$, and decay reactions at rates $\mu, \nu$ that transform $V$ and $U$ into inert products $P$ and $Q$. Finally, $U$ is fed into the system at a rate $f$ and both $U$ and $V$ are allowed to diffuse with diffusion constants $D_u$ and $D_v$ respectively.  We determined the \emph{form} of the intrinsic noise associated with this system of reactions through a procedure which took us from the Master equation describing its chemistry (and that captures its Markovian nature) to an effective non--equilibrium field theory action (Cf. Appendix.~\ref{sec:master}). This  enabled us to derive a set of Langevin equations that incorporated the effects of the intrinsic noise. The structure of the noise was described through unique correlation functions. These  required the existence of a collective mode $\sigma = \lambda_0 \phi_u \phi_v ^2$ where the $\phi_i$ are the fields encoding the chemical concentrations of the species. 

In this sequel paper we focus on the \emph{effects} of this noise.  Specifically we seek to determine whether the physical parameters of the model inherit a scale-dependence as a function of the noise, and if so, whether this induces \emph{new} interactions (relevant and irrelevant) apart from those initially present in the macroscopic chemistry specified by Eq.~\eqref{reactions}. We answer this question positively and consequently investigate the spatiotemporal scales at which these induced interactions manifest themselves along with the momentum space behavior of the parameters and chemical concentrations in the limiting regimes. To denote the limits we employ the field theory ``jargon" whereby the large spatiotemporal scales corresponding to low wave-number and  frequency are referred to collectively as the infrared (IR)  regime, whereas the small spatiotemporal scales corresponding to high wave-number and  frequency is the ultraviolet (UV)  regime. 

Since our goal is to determine the scaling behavior between the two limits, a natural way to proceed is to use a renormalization group approach~\cite{Lee_1994,THVL_2005} on the many-body description of~\eqref{reactions} that we studied in~\cite{CGPM_2013}. Interestingly, we find that the irreversibility of the reactions leads to a time-directionality associated with the Feynman diagrams describing the interactions. This in turn \emph{severely restricts} the possible topologies of the graphs and allows us to carry out a systematic \emph{exact} calculation to determine the effect of the noise.

We find that the strength of the coupling $\lambda_0$ that regulates the autocatalytic part of the chemical reaction is renormalized due to the Markovian nature of the process. Specifically the chemically relevant inelastic collisions  ($U+2V \rightarrow 3V$) proceed through an arbitrary intermediate number of elastic collisions ($U+2V \rightarrow U+2V$) modifying the coupling strength as we change scales. This scale dependence manifests itself only up to a critical dimension $d_c = 1$, above which (in the absence of cutoffs) the coupling constant is formally zero in the IR. Thus beyond $d_c=1$ the system must be considered an \emph{effective field theory}, whose parameters must be determined by low momentum experiments.  The number of parameters of this effective theory are most easily described in terms of recasting it via composite fields for the concentrations $\phi_u \phi_v ^2$, $\phi_u \phi_v$
and $\phi_v ^2$.  In particular the reaction $U+2V \rightarrow 3 V$ is then interpreted as proceeding through the formation of an intermediate state $\sigma = \lambda_0 \phi_u \phi_v ^2$ that was also essential in determining the Langevin equations for~\eqref{reactions} that were derived in~\cite{CGPM_2013}.

When probing the system at short spatiotemporal (UV) scales it turns out that there are two separate manifestations of $\sigma$ (i) a composite bound state or (ii) an elementary particle (molecule)  with its own bare kinetic terms each leading to \emph{different} equations of motion. However, starting from $d =2$ both versions of $\sigma$ leads to the same equations of motion. 
The physical implication of this is that one cannot experimentally resolve $\sigma$ into its constituents and that it behaves as an elementary particle. In order to see its composite nature one would need to probe at spatio-temporal scales shorter than that associated with the mean free path of the chemicals.  However in this limit the Markovian assumption is violated and new physics is required to describe the chemistry. 

The fluctuations also lead to relevant new noise-induced interactions of the form $n \rightarrow n$ for $n \ge 4$ where $n$ is the number of molecules entering or leaving the reaction zone.  Although all these higher order processes are naively divergent, to regulate them, one might think that an infinite number of ``counter terms" need to be added.  In two dimensions, however, it turns out that by again introducing composite concentration fields for the di-molecules  $\phi_u \phi_v$ and $\phi_v ^2$, the divergences in these induced interactions can be regulated by introducing  just two new effective field theory parameters: the decay rates (masses) $M_{1,2}$ for $\phi_u \phi_v$ and $\phi_v ^2$. 
Finally in three dimensions no new parameters are needed, however the equations of motion are modified: in order  to describe the infrared physical chemistry at small momentum one needs to introduce higher order kinetic terms into the action for $\sigma$ .

\section {Momentum-dependent reaction rate}

\begin{figure}[t!]
\centering 
\includegraphics[width=0.5\textwidth]{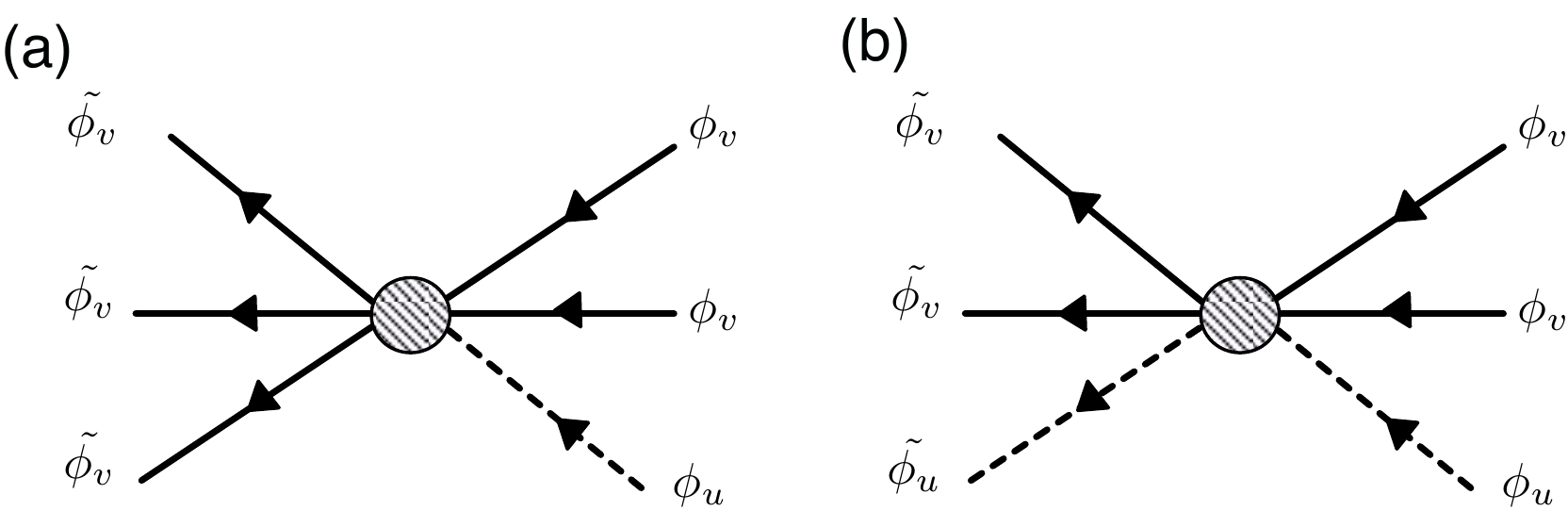}
\caption{Many body interpretation of the reaction $U+2V \rightarrow 3V$. (a) Inelastic scattering where one molecule of $U$ and two molecules of $V$ are destroyed at an interaction vertex (shaded) to create three molecules of $V$. (b) The elastic scattering where $U+2V \rightarrow U+2V$ which must be present in order for particle number and probability conservation. Time flows from right to left.}
\label{fig:fig1}
\end{figure}

In order to uncover the momentum scale behavior of the constituents of the model, we will resort to the many body description of Eq.~\eqref{reactions}. In this description the reaction $U+2 V \rightarrow 3V$ is interpreted in terms of a three body inelastic collision where two particles of $V$ and one of $U$ are destroyed at an interaction vertex to create three of $V$.  When one writes down the master equation describing this reaction (see Eq.~\eqref{master2}), one finds that in order to conserve probabilities, it is also necessary to include the elastic collision $U+2V \rightarrow U+ 2V$.  A graphical representation of this is shown in  in Fig.~\ref{fig:fig1}, where the inelastic scattering is proportional to $-\lambda_0$, while the elastic scattering is proportional to $\lambda_0$.  Following this, the Doi-Peliti operator technique~\cite{Doi76, Grassberger_Scheunert} is used to write down an equivalent non-equilibrium field theory action~\cite{NO98, Tauber_2007,ZHM06} thus,
\ba
S = && \int dx \int_0^\tau dt \bigl[ \phi_v^\star \partial_t \phi_v + D_v \nabla \phi_v^\star \nabla \phi_v  +\phi_u^\star \partial_t \phi_u \nonumber \\
&&+ D_u \nabla \phi_u^\star \nabla \phi_u+ \mu (\phi_v^\star-1)  \phi_v + \nu (\phi_u^\star -1) \phi_u \nonumber \\
&& - f (\phi_u^\star -1)  
 - \lambda_0(\phi_v^\star-\phi_u^\star)\phi_v^{\star 2}  \phi_v^2 \phi_u  \bigr],
 \label{actionunshifted}
\ea
where the $\phi_i$'s are fields representing the concentrations of the chemicals. (An outline of this method is shown in Appendix~\ref{sec:master}, see~\cite{CGPM_2013} for more extensive details of the derivation.)

\begin{figure*}[t!]
\centering 
\includegraphics[width=0.9\textwidth]{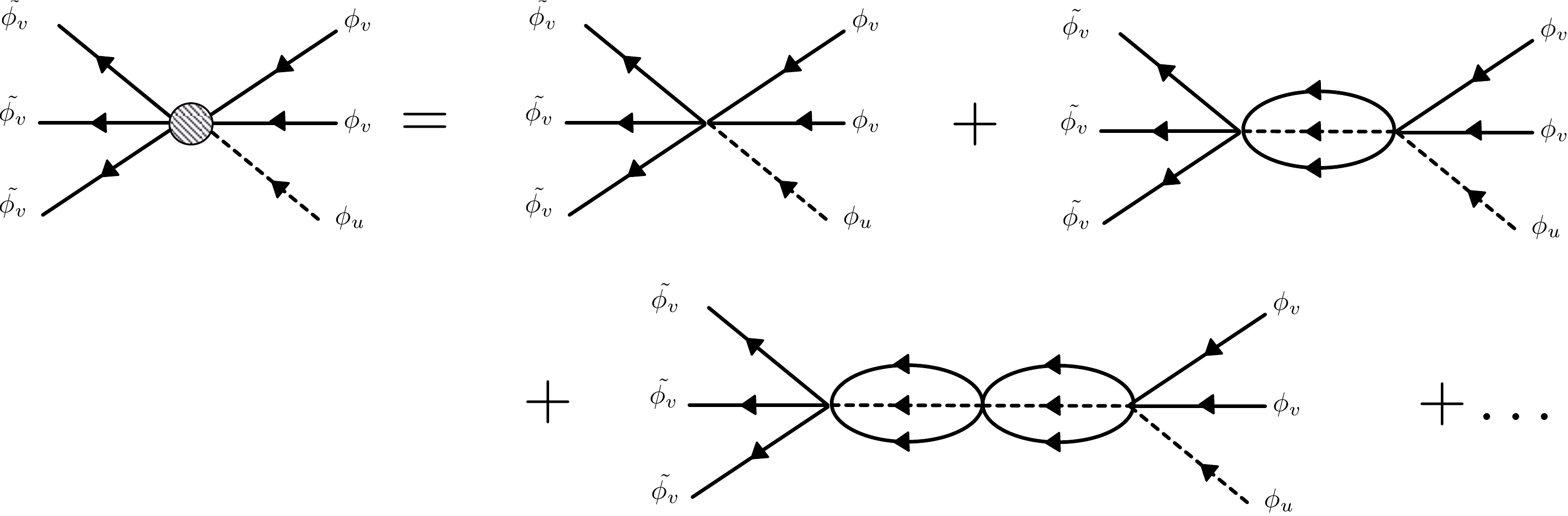}
\caption{The diagrammatic expansion for the vertex function $\Gamma^{(3,3)}$ for inelastic scattering. The same set of primitive divergences contribute to that for elastic scattering, which is the same except for the substitution of a single outgoing leg from $\phi_v$ to $\phi_u$. Consequently both processes are renormalized through a single renormalized coupling constant.}
\label{fig:fig2}
\end{figure*}

Since the action must be dimensionless,  if we introduce a momentum scale $[q] = \kappa$ along with the diffusive temporal scaling $[t] = \kappa^{-2}$,  it is straightforward to see that we have associated scaling dimensions:
\bq
[\phi_{v,u}] = \kappa^d, \quad [\phi_{v,u}^\star] = \kappa^{0}, \quad [\lambda_0] = \kappa^{2\epsilon/d_c},
\label{scaling}
\eq
where the starred fields are chosen to be dimensionless by convention and
$\epsilon = d_c - d$,
where $d_c$ is the critical dimension (i.e. the dimension below which $\lambda_0$ has non-vanishing finite value in the high-momentum UV limit). Generalizing the cubic interaction $ \phi_v^2 \phi_u$ in~\eqref{actionunshifted} to the form $\phi_v ^{k} \phi_u ^{\l}$, the general form of the critical dimension is seen to be 
$d_c(k,l) = 2/(k+l-1)$ which in this case (where $k = 2, l=1$) implies that $d_c = 1$.    Consequently the dimensionless version of the coupling constant $\lambda_0$ is therefore the combination $\lambda_0  \kappa^{-2 (1-d)}$.  

It is worth noting that the action shown in~\eqref{actionunshifted} corresponds to a ``symmetric" phase in the sense that it possesses a $U(1)$ symmetry due to particle number conservation (essentially equivalent to a form of the classic Lavoisier's principle).  The graphs that contribute to $any$ one particle irreducible vertex process or any intermediate state, must be constructed from the (directed) basic interactions shown in Fig.~\ref{fig:fig1}.  
The $U(1)$ symmetry of the basic $3 \rightarrow 3$ process corresponding to the inelastic chemical reaction $U+2V \rightarrow 3V$ (as well as the corresponding elastic reaction $U+2V \rightarrow 3V$), along with the unidirectionality of the reactions (being non-reversible) severely restrict the set of graphs that can be constructed.  In fact looking at the topological structure of the bare vertices in Fig.~\ref{fig:fig1}, it becomes clear that there is no combination that can generate any diagram contributing to corrections to the propagator ($\phi_i \leftrightarrow \phi_i$), since this requires (graphically) one incoming and one outgoing line.  This implies in turn that there is no momentum dependence of the diffusion constants $D_u, D_v$ or the decay terms $\mu, \nu$. 

On the other hand, as shown in Fig.~\ref{fig:fig2}, it is clearly possible to rewrite the bare three body interaction in terms of a diagrammatic expansion corresponding to a perturbation series in $\lambda_0$.  Each component in the expansion corresponds to an elastic re-scattering represented through a loop $I_2(p,t)$ (the subscript 2 reflects the fact that the elastic   re-scattering consists of two loops).  Mathematically this is expressed through the vertex function $\Gamma_{\textrm{i}}^{(3,3)}$ (where the 3 refers to the number of incoming and outgoing lines in Fig.~\ref{fig:fig1}) whose formal expression is
\ba
\Gamma^{(3,3)}_{\textrm{i}} (p,t)&= \lambda_0 \delta(t_2-t_1) - \lambda_0 ^2 I_2(p, t_2-t_1) +\lambda_0^{3} \nonumber \\
&\times \int_{t_1}^{t_2} dt' I_2(p,t_2 - t')I_2(p,t'-t_1) - \ldots,
\label{vertex}
\ea
where the subscript i refers to inelastic. (The vertex function for elastic scattering $\Gamma_{\textrm{e}}$ is just -$\Gamma_{\textrm {i}}$.) 

Assuming there is an external momentum $p$ flowing into the graph from the three external legs, the explicit expression for the loop $I_2(p,t)$ is,
\ba 
I_2(p,t) = &&3! \int \prod_{i=1}^{3} \left(\frac{d^{d} p_i}{(2 \pi)^{d}}\right) e^{-[D_v(p_1^2+p_2^2)+ D_u p_3^2 + 2 \mu + \nu]t} \nonumber \\
&& \times(2 \pi)^d \delta(p-p_1-p_2-p_3).
\ea
Through appropriate linear transformations and (legal) shifts of the momentum variables, it is easily shown that
\bq
I_2(p,t) =B_2(d)  t^{-d} \exp\left[-\left(\alpha_2 + \tilde{D}_2 p^2\right) t \right],
\eq
where to avoid clutter we use the abbreviations $B_2(d)  =  \frac{3!}{(4 \pi D_2)^{d}}, \quad D_2 = \sqrt{D_v (2D_u+D_v)}, \quad \tilde{D}_2 = \frac{D_u D_v ^2} {D_2 ^2}$ and $\alpha_2 =   2 \mu + \nu$.
Taking the Laplace transform of \eqref{vertex} renders it into a geometric sum via the convolution theorem and yields,
\ba
\Gamma^{(3,3)}_{\textrm{i}} (\tilde s) &=& \frac{\lambda_0}{1+\lambda_0  I_2(\tilde s)}, \nonumber \\
I_2(\tilde s) &=&  B_2(d)  \Gamma(1-d) \left(\tilde s + \alpha_2\right)^{-(1-d)},
\label{I2}
\ea
where $\Gamma(x)$ is the Euler-Gamma function and $\tilde s =s + \tilde{D}_2 p^2$. 
From~\eqref{I2} it is clear that $\lambda_0$ is the reaction rate in the {\it absence} of fluctuations (since in that case $I_2(\tilde s) = 0)$, whereas the full expression $\Gamma_{\textrm{i}} ^{(3,3)}$ reflects its  function on the energy and momentum scale at which it is measured. It is therefore interpreted as the \emph{dimension-full running} reaction rate.  

The fluctuation term $I_2(\tilde s)$ has a pole in one dimension and in \emph{every} positive integer dimension.  For physical reasons we restrict ourselves to the range $1 \le d \le 3$ and therefore we expose the relevant divergences through the identity $\Gamma(n+1) = n \Gamma(n)$, to rewrite 
\bq
\Gamma(1-d) = \frac{\Gamma(4-d)}{(1-d)(2-d)(3-d)}.
\eq
In order to carry out the corresponding renormalization---as in standard quantum field theory---it is convenient to introduce a dimensionless counterpart to $\lambda_0$.  As mentioned before the dimensionless \emph{bare} reaction rate in the presence of a momentum scale $\kappa$ is 
$\lambda_0 \kappa^{-2(1-d)} $ . However, in order to simplify the algebra we will instead find it convenient to work with a slightly modified form thus
\bq
g_0(\kappa) = \lambda_0 \kappa^{-2(1-d)} B_2(d) \Gamma (4-d)  \label{bare}
\eq
in terms of which the dimensionless running reaction rate is modified to
\ba 
g_R (\tilde s) &&={\Gamma}^{(3,3)}_{\textrm{i}}(p,s) B_2(d)  \Gamma (4-d) \kappa^{-2(1-d)} \nonumber \\
&& =\frac{g_0(\kappa)}{1+ \frac{g_0(\kappa)}{(1-d)(2-d)(3-d)} \left(\frac{ \tilde s+ \alpha_2 }{\kappa^2}\right)^{-(1-d)}}.
\label{grps1}
\ea
By itself, this does not seem particularly useful, since we need to translate this into a physically measurable quantity. To do so we begin by defining a renormalized coupling constant  $g_R (\kappa)$ at the convenient renormalization point ${\tilde s_0}  =\kappa^2 -\alpha_2$.

\begin{figure*}[t!]
\centering 
\includegraphics[width= \textwidth]{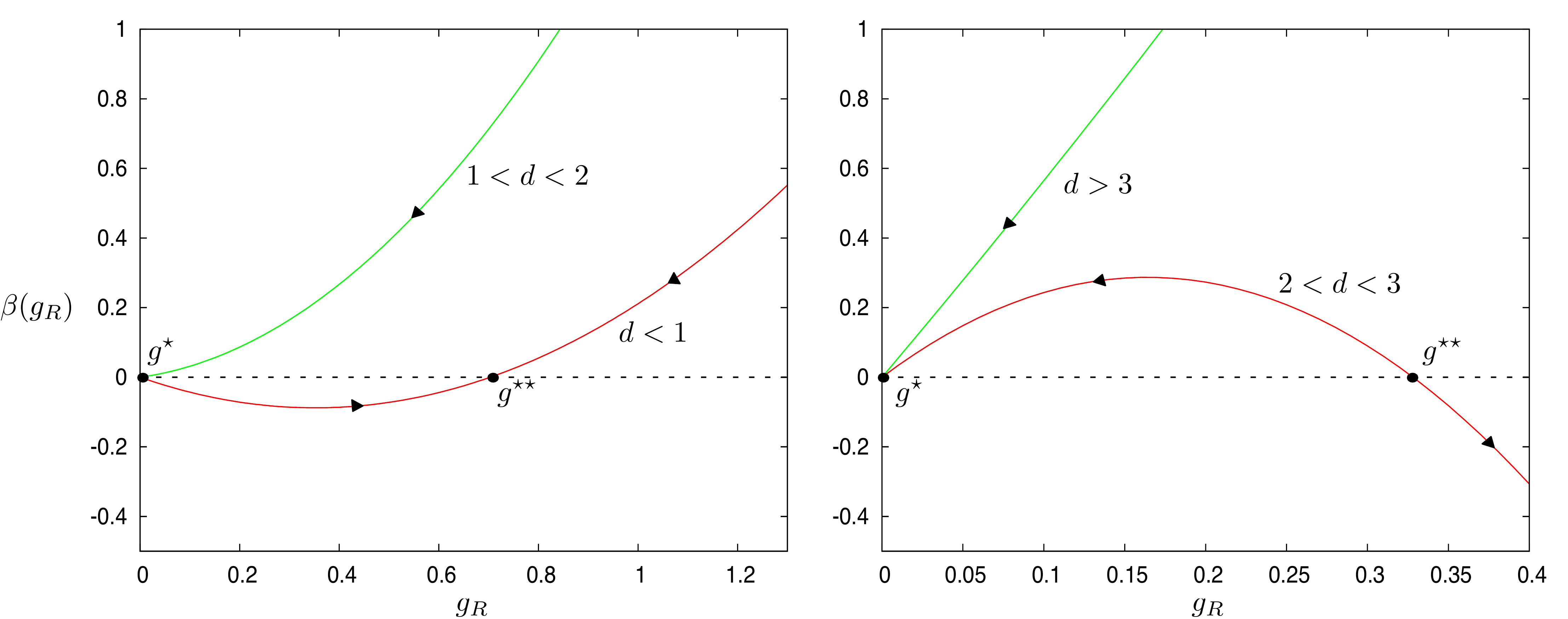}
\caption{The Beta function~\eqref{beta} as a function of the dimensionless coupling $g_R ({\tilde s_0})$. Note that this refers to IR and \emph{not} UV flow. There are two fixed points in the regions $d < 1$ and $2 < d < 3$, whereas there is a single fixed point when $1 < d < 2 $ and $d > 3$. }
\label{fig:fig3}
\end{figure*}
 
Physically, this implies that we measure (experimentally) the coupling at $\tilde s_0$ and use it to determine what value of the bare dimensionless coupling $g_0$ corresponds to the \emph{physically measured} running $g_R$.  
Of course, once this measurement is made, the value of the coupling at any other momentum scale is determined. 

For this choice of measurement scale the connection between $g_R$ and $g_0$ simplifies to  
\bq
g_R( {\tilde s_0})  =  \frac{g_0(\kappa) }{1+ \frac{g_0(\kappa) }{(1-d)(2-d)(3-d)} }. 
\label{gr}
\eq
with $g_0(\kappa)$ given by Eq. \eqref{bare}. Finally, combining Eqns.~\eqref{grps1} and~\eqref{gr}, the expression for the (running) coupling constant for arbitrary momentum and energy scale is 
\ba
g_R(\tilde s) &=& \frac{g_R( {\tilde s_0})}{1+\frac{g_R( {\tilde s_0})}{(1-d)(2-d) (3-d)}\left( \left(\frac{\tilde s + \alpha_2 }{\kappa^2}\right)^{-(1-d)}-1\right)},
\nonumber \\ \label{run}
\ea
and of course by inspection, it is apparent that at the renormalization point  $\tilde{s}_0 = \kappa^2 - \alpha_2$  one obtains the equivalence $g_R (\tilde s_0) = g_R (\tilde s_0)$ as should be the case by definition. 

We see that for $d<1$ the coupling in Eq. (\ref{grps1}) is finite in both the UV and IR regimes and as $\tilde s \rightarrow \infty$ we get $g_R \rightarrow g_0$. 
At the critical dimension $d_c = 1$ we can expand the denominator in a power-series to get
\bq
g_R(\tilde s) = \frac{g_R( {\tilde s_0})}  { 1- \frac{ g_R( {\tilde s_0})} {2}  \ln  { \left(\frac{\tilde s + \alpha_2 }{\kappa^2}\right)}},
\label{running}
\eq
which suggests that there is a Landau pole at $g_R( {\tilde s_0}) = 2 / \ln  { \left(\frac {\tilde s + \alpha_2 }{\kappa^2}\right)}) $.  A similar situation is found in quantum electrodynamics where such a singularity is what leads to the divergence of the bare charge $e_0$ in the UV limit. Correspondingly, this is interpreted as the harbinger of $new$ phenomena or degrees of freedom in the UV or, equivalently, at the shorter length scales, and which in a chemical context hints at the presence of short lived or intermediate substances in the mechanism of the original chemical reaction.

A naive use of this dimensionally regulated answer yields as $d \rightarrow 2$ the result
\bq
g_R(\tilde s) = \frac{ (2-d)  g_R( {\tilde s_0})}{(2-d)-  g_R( {\tilde s_0}) \left( \frac{\tilde s +\alpha_2}{\kappa^2} -1\right)},
\label{d-2}
\eq
implying that the coupling goes to zero in the continuum limit.  
This can also be seen through the associated $\beta$ function for $g_R ( {\tilde s_0})$ which has the particularly simple form,
\ba
\beta(g_R( {\tilde s_0})) &=& \kappa \frac{dg_R( {\tilde s_0})}{d \kappa} \nonumber \\
&=& 2 g_R( {\tilde s_0}) \left( d-1+ \frac{g_R( {\tilde s_0})}{(2-d)(3-d)} \right).
\label{beta}
\ea
This has two fixed points:  a trivial one at $g_R ^{\star} = 0$ and a non-trivial one at $g_R ^{\star \star} = (1-d) (2-d) (3-d) $. Note that the non-trivial fixed point exists \emph{only} when $ d < 1$ or $2< d< 3$. In the former case $g_R^{\star \star}$ is IR stable while $g_R^{\star}$ is UV stable, with the situation being reversed in the latter case. On the other hand when $1 < d < 2$ and $d >3$ there is only a single fixed point $g_{R}^{\star}$ which is IR stable but UV divergent (see Fig.~\ref{fig:fig3}). 

Taken naively this seems to indicate that there is no momentum dependence of the coupling once we are beyond one dimension. In reality, however, there is a characteristic length scale in the system.  Indeed, as discussed in the introduction, beneath the development of the master equation and its corresponding action~\eqref{actionunshifted} lurks the assumption of a lattice on which the chemicals hop between sites. In addition, the size of the lattice must be larger than the mean free path of each chemical species in order to preserve the  $assumed$ Markovian nature of the collisions. In other words the term $(d-2)$ in~\eqref{d-2} really indicates a cut-off of the form $1/ \ln (\Lambda^2)$, where $\Lambda$ is the maximum limit of the lattice momentum.  Therefore in two dimensions, the action~\eqref{actionunshifted} represents an \emph{effective} field theory with an UV cutoff. 

In three dimensions we get the dimensionally regulated answer:
\bq
g_R(\tilde s) = \frac{ (3-d)  g_R( {\tilde s_0})}{(3-d)+  \frac{1}{2} g_R( {\tilde s_0}) \left( \left( \frac{\tilde s +\alpha_2} {\kappa^2}\right)^2 -1\right)},
\label{d-3}
\eq
which suffers from the same pathologies as in the two-dimensional case.

In order to obtain the correct effective theory, we make use of auxiliary fields that allows us to interpret the sum of loops in the vertex function $\Gamma^{(3,3)}_{\textrm{i}}$ (i.e. the sum of elastic scatterings) as going through a single composite state $\sigma$. We can imagine the field $\sigma$ as representing the density of a ``cloud" of chemicals involved in the elastic scattering. This field then requires a ``mass" (decay rate) and  wave-function renormalization to render the system finite. The effective theory can then be described in terms of some low energy parameters such as long distance reaction rates, as well as the low momentum decay rates of composite fields. As is well known, this can be made explicit by making use of the well known technique of the Hubbard-Stratonovich \cite{Strat1,r:Hubbard:1959kx}  transformation, which we describe and apply next to this problem.
 
 \begin{figure*}[t!] 
\centering 
\includegraphics[width=0.95\textwidth]{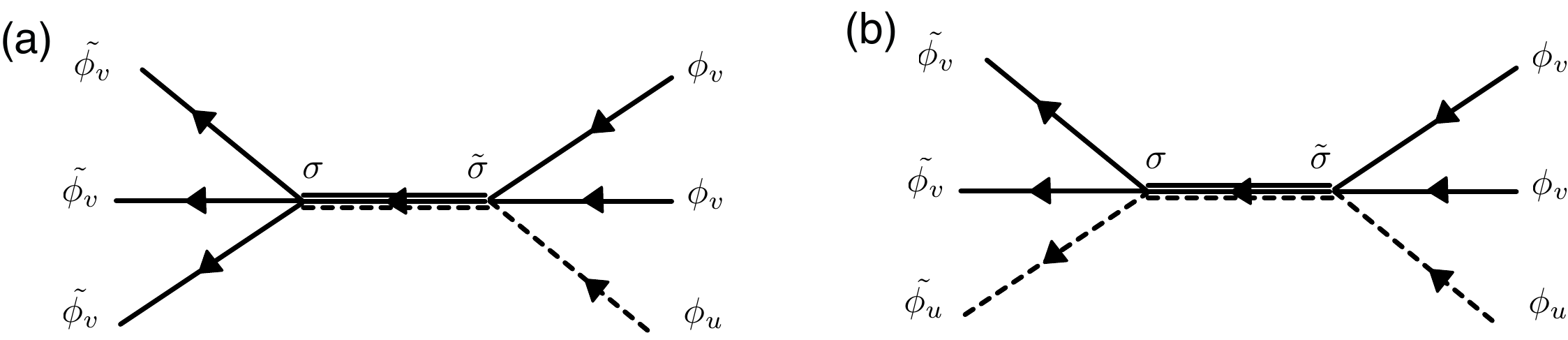}
\caption{The vertex function $\Gamma_{\textrm i}^{(3,3)}$ shown in  Fig.  \ref{fig:fig2} reinterpreted as the production and decay of the composite field $\sigma$.  The composite field $\sigma$ is  formed out of the combination $\lambda \phi_u \phi_v ^2$ and leads to the  two decay channels shown in  (a) and (b).}
\label{fig:fig5}
\end{figure*}

\section{Composite field operators}

We now introduce the composite fields $\sigma$ and $\sigma^\star$ into the previous field theory via a Hubbard-Stratonovich transformation and discuss the 
differences between an ``elementary" $ \sigma$ field and a composite one.  We will find that at large times and distances,  these two theories  are indistinguishable. However at short scales (high momentum), these theories differ for $d =1$.

Our starting point, once again, is the unshifted action~\eqref{actionunshifted}
\ba
S = && \int dx \int_0^\tau dt \bigl[ \phi_v^\star \partial_t \phi_v + D_v \nabla \phi_v^\star \nabla \phi_v  +\phi_u^\star \partial_t \phi_u \nonumber \\
&&+ D_u \nabla \phi_u^\star \nabla \phi_u+ \mu (\phi_v^\star-1)  \phi_v + \nu (\phi_u^\star -1) \phi_u \nonumber \\
&& - f (\phi_u^\star -1)  
 - \lambda_0(\phi_v^\star-\phi_u^\star)\phi_v^{\star 2}  \phi_v^2 \phi_u  \bigr].
 \label{actionunshifteda}
\ea
To carry out the Hubbard-Stratonovich transformation, we construct a second action which defines the composite fields $\sigma$ and $\sigma^\star$ thus,
\ba
S' =  \int dx &&\int_0^\tau dt \lambda_0  \bigl[  \left( (\sigma^\star-1) - (\phi_v^\star-\phi_u^\star)\phi_v^{\star 2} \right)\nonumber \\
&&\times (\sigma -   \phi_v^2 \phi_u  ) \bigr], 
\ea
and add it to $S$ obtaining,
\ba
S_{\textrm{comp}} &=& S + S'  \nonumber \\
&= & \int dx \int_0^\tau dt \bigl[ \phi_v^\star \partial_t \phi_v + D_v \nabla \phi_v^\star \nabla \phi_v  +\phi_u^\star \partial_t \phi_u, \nonumber \\
&&+ D_u \nabla \phi_u^\star \nabla \phi_u+ \mu (\phi_v^\star-1)  \phi_v + \nu (\phi_u^\star -1) \phi_u \nonumber \\
&& - f (\phi_u^\star -1)  - \lambda_0 \sigma (\phi_v^\star-\phi_u^\star)\phi_v^{\star 2}  \nonumber \\
 && -  \lambda_0(\sigma^\star -1)   \phi_v^2 \phi_u +\lambda_0 (\sigma^\star-1) \sigma \bigr]. 
 \label{action2}
\ea
Eliminating the constraint equation defining $\sigma$ leads back to the original equation of motion for the fields $\phi_i$. From \eqref{action2}, it is apparent that
while the composite field $\sigma$ is formed through a combination of a single $\phi_u$ and two $\phi_v$'s, it has two potential decay channels (i) it converts into either 3 $\phi_v$'s or  (ii) back again to the original constituents, a single $\phi_u$ and two $\phi_v$'s (Cf. Fig.~\ref{fig:fig5}).  

It is instructive to perform a comparison of the above, with the case where we consider $\sigma$ as an \emph{elementary} scalar particle.   In this version,  instead of the local interaction $\lambda_0(\sigma^\star -1) \sigma$ found in  $S_{\textrm{comp}}$ there is a fundamental (or rather bare) kinetic energy term for the $\sigma$ field, as well as an unrenormalized decay rate $M_0$. To keep the dimensionality of the elementary $\sigma$ the same as in the composite case, the ``free" part of the $\sigma$ field action needs to be divided by $\kappa^{2d}$.   Consequently, the action is now

\ba
S_{\textrm{elem}} = && \int dx \int_0^\tau dt \bigl[ \phi_v^\star \partial_t \phi_v + D_v \nabla \phi_v^\star \nabla \phi_v  +\phi_u^\star \partial_t \phi_u \nonumber \\
&&+ D_u \nabla \phi_u^\star \nabla \phi_u+\left( \sigma^\star \partial_t \sigma 
+ D_\sigma  \nabla \sigma^\star \nabla \sigma \right) \kappa^{-2d} \nonumber  \\
&& + 
 \mu (\phi_v^\star-1)  \phi_v + \nu (\phi_u^\star -1) \phi_u - f (\phi_u^\star -1)             \nonumber \\
&& + M_0 \kappa^{-2d} (\sigma^\star-1) \sigma - \lambda_0 \sigma (\phi_v^\star-\phi_u^\star)\phi_v^{\star 2}  \nonumber \\
 &&- \lambda_0(\sigma^\star -1)   \phi_v^2 \phi_u \bigr].
 \label{action4}
\ea
Note that in the models defined by either $\eqref{action2}$ or $\eqref{action4}$, the structure of the Schwinger-Dyson equations are quite similar. 

Having introduced $\sigma$, the process of elastic scattering is now interpreted as proceeding through the exchange of a composite field. This involves the composite propagator, whose inverse is given by  the dimensionally regulated expression
\ba
G_{\textrm{comp}} ^{-1} ({\tilde s})  =   g_0(\kappa) &+& \frac{g_0^2(\kappa)}{(1-d)(2-d)(3-d) } \nonumber \\
&\times& \left(\frac{{\tilde s}+ \alpha_2 }{\kappa^2}\right)^{-(1-d)} \,.
\label{Gcomp}
\ea
In the case where $\sigma$ is an elementary particle, this is instead 
\ba
G_{\textrm{elem}} ^{-1}({\tilde s})   =  \frac{{\tilde s}+ M_0} {\kappa^2} &+& \frac{g_0^2(\kappa)}{(1-d) (2-d)(3-d)}\nonumber \\
&\times& \left(\frac{{\tilde s}+ \alpha_2 }{\kappa^2}\right)^{-(1-d)},
\label{Gelem}
\ea
where for the sake of simplicity we have chosen the same diffusion constant for the free part as in the induced one, i.e. $D_\sigma = \tilde{D}_2$. 
%The vertex for $ \sigma \rightarrow \phi_u \phi_v^2$  in dimensionless form  is given by $g_0(\kappa)$. 
% We recognize that we can write the running coupling constant $g_R({\tilde s}) $ in the form consistent with the new vertex and composite field propagator. Namely we have 
%\bq
%g_R ( {\tilde s}) \equiv  g_0^2(\kappa) G_{\textrm{comp}} ({\tilde s}) .
%\label{GML}
%\eq
%

To complete the renormalization procedure in terms of the composite field  $\sigma$ 
we must now allow for ``wave function" renormalization for the $\sigma$ field as well as ``mass" (decay rate) renormalization.  Since the vertex function $\Gamma_{\textrm{i}}^{(3,3)}$ depends on the combination $\tilde s = s+ {\tilde D}_2 p^2$, the inverse propagator~\eqref{Gcomp} can therefore be expanded in a power series in $\tilde s$ thus
 \begin{widetext}
\ba
G_{{\rm comp}} ^{-1}&=&  g_0(\kappa)+ g_0^2(\kappa) F(\tilde s) \equiv  g_0(\kappa)+ \frac{g_0^2(\kappa)}{(1-d)(2-d)(3-d)} \left(\frac{\alpha_2 + \tilde{s} }{\kappa^2}\right)^{-(1-d)}   \nonumber \\
&=& 
 g_0(\kappa)+ \frac{g_0^2(\kappa)}{(1-d)(2-d)(3-d)}\left(\frac{\alpha_2}{\kappa^2}\right)^{-(1-d)}  
 \times \nonumber \\
&& \biggl[ 1+\frac{(d-1)  \tilde s}{\alpha_2 }+ \frac{(d-2) (d-1)  \tilde s^2}{2 \alpha_2
   ^2}+ \frac{(d-3) (d-2) (d-1)  \tilde s^3}{6 \alpha_2 ^3}
+ \ldots  \biggr], \nonumber \\  \label{expand}
\ea
\end{widetext}
where $F(\tilde s)$ can be thought of as a self-energy function. The expansion for the elementary $\sigma$ is identical, with the exception that the first term in~\eqref{expand} i.e $g_0(\kappa)$, is replaced by  $ \left(M_0 + {\tilde s}\right)/\kappa^2$. Otherwise the renormalization procedure is identical in both cases. Let us first consider the situation for $d < 3$.

\subsection{One and two dimensions}

For both the composite and elementary $\sigma$'s we can rewrite $G^{-1}$ in the form
\bq
G ^{-1}   = Z_3^{-1} \left[ \frac{ M_\sigma  + \tilde s }{\kappa ^2} + g_0^2(\kappa)Z_3 F_{sub_2} (\tilde s ) \right],
\eq
for $d <3$ and the quantities $Z_3$ and $M_\sigma$ are obtained from the first two terms in the expansion of Eq.~\eqref{expand}. (We will henceforth drop the subscript for $G$ and unless mentioned otherwise it refers to both the elementary and composite versions of the model.)  The subscript $sub_2$ refers to the subtraction of  the first two terms in the power series~\eqref{expand}, $M_\sigma$ is the {\it renormalized} decay rate for the $\sigma$ field and $Z_3$ is the wave function renormalization.  Note that at low momentum, apart from the rescaling by $Z_3^{-1} $, the inverse propagator resembles the free field one.

When $\sigma$ is composite this leads to the identities
\ba
 Z_3^{-1} \frac{ M_\sigma }{\kappa^2}  &=&  g_0(\kappa)+ \frac{g_0^2(\kappa)}{(1-d) (2-d) (3-d)}\left(\frac{\alpha_2}{\kappa^2}\right)^{-(1-d)},  \nonumber \\
 Z_3^{-1} \frac{ \tilde s }{\kappa^2}  &=&   -\frac{g_0^2(\kappa)}{(2-d)(3-d)} \left(\frac{\alpha_2}{\kappa^2}\right)^{-(2-d)}  \frac{ \tilde s }{\kappa^2}. 
 \label{com1}
 \ea
whereas when it is considered elementary we have,
\ba
 Z_3^{-1} \frac{ M_\sigma }{\kappa^2} &=& \frac{ M_0}{\kappa^2} + \frac{g_0^2(\kappa)}{ (1-d) (2-d)(3-d)}\left(\frac{\alpha_2}{\kappa^2}\right)^{-(1-d)}, \nonumber \\
 Z_3^{-1} \frac{ \tilde s }{\kappa^2}  &=&   \left( 1- \frac{g_0^2(\kappa)}{(2-d)(3-d)} \left(\frac{\alpha_2}{\kappa^2}\right)^{-(2-d)}  \right)  \frac{ \tilde s }{\kappa^2}.  
 \label{elem1} 
 \ea
Next we introduce the renormalized vertex $g_\sigma $  for $ \sigma \rightarrow  \phi_u \phi_v^2 $ making use of the identities
\bq
g_\sigma ^2 ={g^2 _0}(\kappa) Z_3; \quad \quad G_R = Z_3^{-1} G,
\label{gsigmagzero}
\eq
where $G_R$ is the renormalized propagator for $\sigma$. Consequently the combination 
\bq
g_0 ^2 ( \kappa) G = g_\sigma ^2  G_R 
\label{identity}
\eq
is invariant under renormalization. Combining these leads us to a finite expression for the running reaction rate thus
\bq
g_R(\tilde s) =  g_\sigma ^2 G_R= \frac {g_\sigma ^2 } {   \frac{ M_\sigma  + \tilde s }{\kappa ^2} + {g_\sigma}^2 F_{sub_2} (\tilde s ) }\,.
\eq
In two dimensions $F_{sub_2}$ is just zero and therefore this reduces to
\bq
g_R(\tilde s)= \frac{ g_\sigma  ^2 \kappa^2}
{M_\sigma + \tilde s}. 
%- \hat{D_2} p^2 + \frac{Z_3 g_0(\kappa) ^2}{(1-d)(2-d)} \left[ \left(\frac{s+ \kappa_2 + \tilde{D} p^2}{\kappa^2}\right)^{d-1} - \left(1 - (1-d)\left(\frac{s+{\hat D}_2 p^2}{\kappa}\right)\right)\right]}, 
\label{gr1} 
\eq
In particular using~\eqref{identity} and the definition of $G_R$ we have that 
\bq
g_R(\tilde s =0) = g_\sigma^2  \frac{\kappa^2} {M_\sigma} = g_0^2 (\kappa)  \frac{\kappa^2}{ Z_3^{-1} M_\sigma}.
\eq
Finally, employing Eqns. \eqref{com1} or \eqref {elem1} allows us to relate $g_R(0) $ to $g_0(\kappa) $.  

We thus conclude that in $d=2$, the renormalized coupling $g_R (\tilde s)$ is proportional to 
$d-2$ through the wave function renormalization term $Z_3$.  This suggests that when we use actual cutoffs to regulate the integrals, the renormalized coupling is related to the inverse of the physical cutoff of the problem.  This confirms our previous intuition that the system can only be described by an \emph{effective} field theory with a momentum space (or spatial) cutoff.  
It is also apparent that there is no difference between a fundamental or composite $\sigma$ at this level in either the IR or UV limits.
The dependence on the parameter $M_\sigma$ can be eliminated by   evaluating the running coupling constant (momentum dependent reaction rate) at a particular reference point $\tilde s_0$.  In terms of this reference point
 $\tilde{s}_0 = \kappa^2 - \alpha_2$  (as chosen for Eq.~\eqref{gr})  one finds the reaction rate is given by
 
 \bq
g_R({\tilde s} ) =  \frac{g_R({\tilde s}_0 )}    { 1+  g_R({\tilde s}_0 ) \frac{ ({\tilde s} - {\tilde s}_0)}{\kappa^2}   }  \,
\eq
thus explicitly showing that $M_\sigma$ can be replaced by the physical quantity  $g_R({\tilde s}_0 )$.

Moving onto one dimension which is the critical dimension, from Eqns. \eqref{com1} and  \eqref{elem1} we find  that the wave function renormalization is finite. Thus only mass renormalization is needed which can again be translated into defining $g_R(\tilde s)$  at a particular reference value 
 $\tilde s_0$.  In this case the second term in~\eqref{expand} is
\bq
g_0^2(\kappa)  F (\tilde s) = \frac{g_0^2(\kappa) }{2 (1-d)} - \frac{g_0^2 (\kappa) }{2} \ln { \frac{\tilde s + \alpha_2}{\kappa^2}},
\eq
which leads to
\bq
G ^{-1}   = Z_3^{-1} \left[ \frac{ M_\sigma  + \tilde s }{\kappa ^2} - \frac{ g_\sigma^2 }{2} \left( \ln \left( \frac{\tilde s + \alpha_2}{\alpha_2}\right)  -\frac{\tilde s}{\alpha_2}   \right) \right].
\eq
Once again making use of~\eqref{gsigmagzero} we find that $g_{\sigma}^2 = -2 \alpha_2/\kappa^2$  and therefore the terms linear in $\tilde s$ cancel.  Thus the renormalized propagator for the composite $\sigma$ is 
\bq
G_R ^{-1}  =  \frac{ M_\sigma }{\kappa ^2} - \frac{ g_\sigma^2 }{2}   \ln \left( 1+ \frac{\tilde s}{\alpha_2}\right),  
\eq
Once again the parameter $M_\sigma$ can be eliminated by defining $g_R(\tilde s)$ at some scale $\tilde s_0$
such that
\bq
g_{R} (\tilde s) = \frac{g_R (\tilde s_0)}{1-  \frac{g_R(\tilde s_0)}{2} \ln
\left(\frac{\tilde s + \alpha_2}{\tilde s_0 + \alpha_2}\right)}
\label{runningcomp}
\eq
%\bq
%\frac{1}{g_R(\tilde s) } - \frac{1}{g_R(\tilde s') }= -\frac{1}{2} \ln \frac{\tilde s + \alpha_2}{ \tilde s' + \alpha_2}  \label{run2}
%\eq
implying that it goes to zero as $(\ln \tilde s)^{-1}$ in the UV limit. Note that this is equivalent to our previous result Eq. \eqref{running}.  

Through a similar sequence of manipulations it can be shown that for the elementary $\sigma$ we have the relation
%For the case of the elementary $\sigma$ we have instead that
%\bq 
%Z_3^{-1} =  1 - \frac{g_0^2(\kappa) \kappa^2}{2 \alpha_2}
%\eq
%So in that case
\bq
g_\sigma^2 = Z_3  g_0^2(\kappa) = \frac{ g_0^2(\kappa)}{1 - \frac{g_0^2(\kappa) \kappa^2}{2 \alpha_2}}
\eq
and therefore this is equivalent to the composite $\sigma$ only when $g_0^2 \rightarrow \infty$ in which case $Z_3 = 0$ (the standard ``compositeness" condition in field theory). 
The corresponding analog to Eq.~\eqref{runningcomp} is
\bq
g_{R} (\tilde s) = \frac{g_R (\tilde s_0)}{1-  \frac{g_R(\tilde s_0)}{2} \ln
\left(\frac{\tilde s + \alpha_2}{\tilde s_0 + \alpha_2}\right)+ g_R(\tilde s_0)  \left(\tilde s - \tilde s_0\right) f(\alpha_2,g_\sigma, \kappa) }
\label{runningelem}
\eq
where $f(\alpha_2,g_\sigma, \kappa) = \frac{2 \alpha_2 + k^2 g_\sigma^2}{2 \alpha_2 k^2 g_\sigma^2}$. 
Here, the running coupling goes to zero \emph{linearly} with $\tilde s$ as opposed to logarithmically. Thus unlike in the two dimensional case there is a difference between the elementary and composite manifestations of $\sigma$. 

\subsection{Higher dimensions $(d \ge 3)$}
The renormalization process  for $d \ge 3$ requires the introduction of higher derivative terms into the effective action for the field $\sigma$.  This is due to the fact that the self-energy function $F( \tilde s)$ in~\eqref{expand} has an increasing number of divergent terms.  

In one and two dimensions, as discussed previously, the first two terms in the series diverge and are regulated via  decay rate and wave function renormalization  respectively. Starting from $d=3$ one begins to induce terms that are not originally present in~Eqns.~\eqref{action2} and~\eqref{action4}. Consequently one needs to subtract three terms from
$F( \tilde s)$ to obtain a finite contribution. (And four in $d=4$ and so on.)  
The resulting renormalized  propagator can be written as 
\bq
	{\tilde g_R}^2 G_R = \frac{ {\tilde g_R}^2 } {   \frac{ M_\sigma  + \tilde s }{\kappa ^2}  +  \frac{1}{2}  \frac {\tilde s^2} {\kappa^4} + {\tilde g_r}^2 F_{sub_3} (\tilde s ) },
	\eq
where the term $ \frac{1}{2}  \frac {\tilde s^2} {\kappa^4} $ corresponds to the addition of a  \emph{new induced} term that needs to be inserted in~\eqref{action2} and~\eqref{action4}. This term has the form
\bq
S_{{\rm ind}} = \int dt \int dx \left[  \sigma^\star(x,t)  \left( \partial_t - D_\sigma \nabla^2 \right) ^2 \sigma (x,t)  \right] \kappa^{-2(d+1)}.
\eq

 \begin{figure*}[t!]
\centering 
\includegraphics[width=0.85\textwidth]{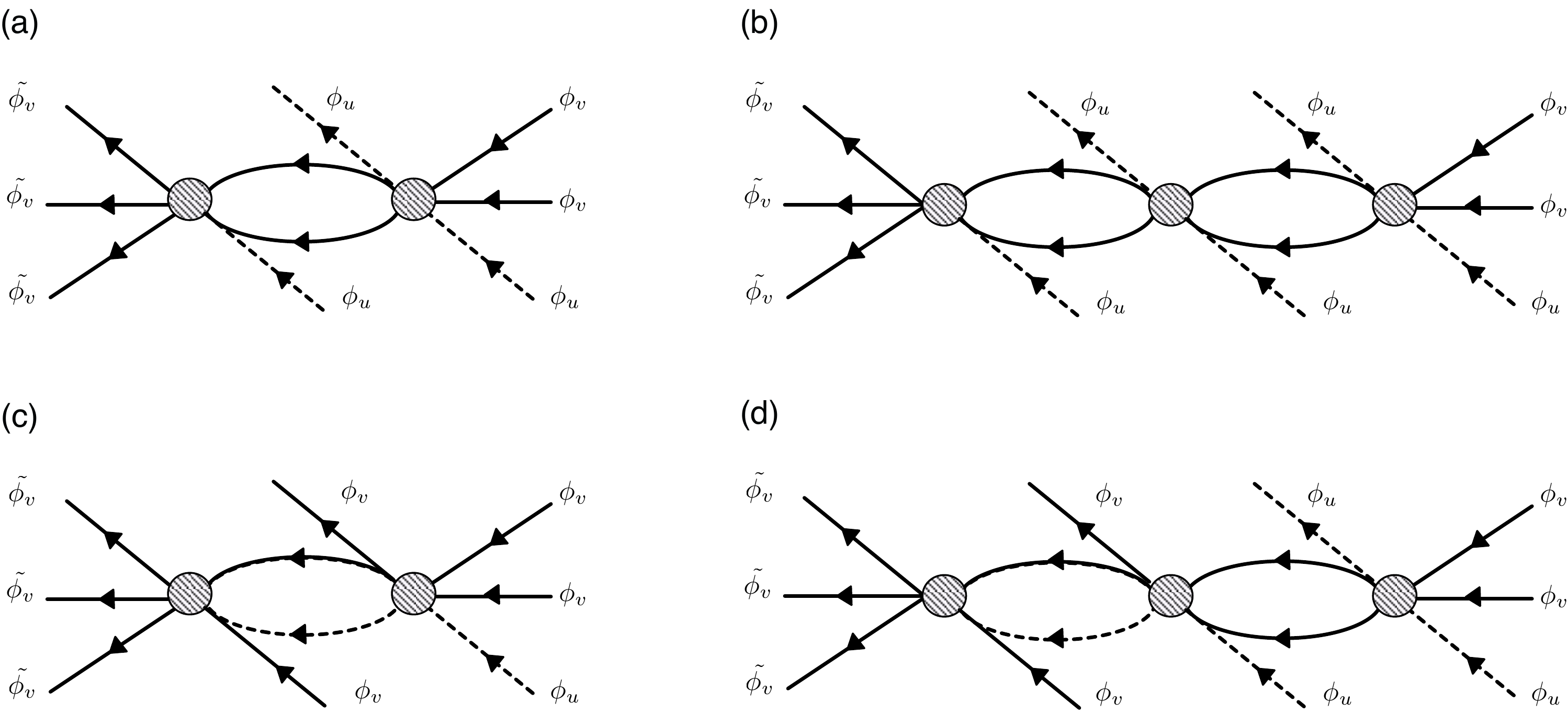}
\caption{Fluctuation induced $n\rightarrow n$ scattering for (a) $n = 4;~~ 2U+ 2V \rightarrow U+ 3V $ and (b) $n=5; ~~3U + 2 V \rightarrow 2 U + 3V$. The shaded circle represents the renormalized original $3\rightarrow3$ scattering from Fig.~\ref{fig:fig2}. These processes proceed through  the single loop consisting of two $\phi_v$'s. which can be thought of as a composite field correlation function for  $\psi_1= \phi_v^2$. We can have related scattering processes  through the composite field correlation function for  $\psi_2 = \phi_{u} \phi_v$  as shown in (c) for the reaction $U + 3 V \rightarrow 4 V$ and indeed a combination of the two composite field correlation functions as shown in (d) for the reaction $2U + 3 V \rightarrow U + 4 V$  .}
\label{fig:fig4}
\end{figure*}

 \section{Fluctuation Induced Processes}

The  interactions shown in Fig. (\ref{fig:fig1})  are such that they can induce processes (via fluctuations) that are not explicitly present in the original  action described by Eq.  \eqref{actionunshifted} .  (For examples of this phenomenon in out of equilibrium RD systems, but generated by $extrinsic$ noise, cf. Ref. {\cite{HLMPM03}.) While this is not problematic if the induced terms are $finite$, a serious problem appears if the terms bring with them a divergence. If each order in perturbation theory leads to new interactions which are divergent, then usually an infinite number of parameters are needed to define the theory.  In such a situation physical predictions cannot be made as they depend on an arbitrary number of parameters.  In that case the theory is called non-renormalizable.  Although we will find that indeed there are an infinite number of 
induced processes that appear superficially divergent, by an appropriate introduction of composite field operators,  we will find that only  three long distance  (infrared) parameters will be  needed to describe all the induced processes. 

The induced processes in this case can be constructed from diagrams that represent $n \rightarrow n$ scattering for $n \ge 4$ for both elastic and inelastic processes. An example of this for $n=4,5$ for the inelastic case is given in Fig.~\ref{fig:fig4} where we consider the diagrams for $2U+ 2V \rightarrow U+ 3V$  in (a)  and  $3U + 2 V \rightarrow 2 U + 3V $ in (b).
Although at first sight it appears  these diagrams are naively divergent,
if one thinks of them as proceeding through ``tree level'' diagrams in terms of the production of di-molecule and tri-molecule states, then
the only  renormalizations necessary are those of the tri-molecule $\sigma$ propagator as well as  decay constant renormalization for the di-molecule correlation functions.  In terms of the renormalized parameters of the $\sigma$ propagator and di-molecule propagators, these
induced processes are then rendered finite. 

Going back to Fig.~\ref{fig:fig4}a,b, we see the formation of a single loop which now consists of two $\phi_v$'s (or, equivalently, a correlation function for the composite field $\psi_1 = \phi_v^2$).   If we instead considered the process $U+3V \rightarrow 4 V$ (c)  this would have been facilitated instead by the correlation function for the composite field $\psi_2= \phi_u \phi_v$.  In terms of the composite fields, these processes are ``tree graphs" in the Green's functions for $\sigma$, $\psi_1$ and $\psi_2$.  
These internal processes are reversible in that    $\phi_u +   \phi_v \rightarrow  \psi_2 \rightarrow \phi_u +   \phi_v$ and $2 \phi_v \rightarrow  \psi_1 \rightarrow  2 \phi_v$ etc.  Starting at $n=5$ we can have both the $\psi_1$ and $\psi_2$ correlation functions occurring in the composite field ``tree" diagram as shown in (d) for the process $2U + 3 V \rightarrow U + 4 V$. 

To expose and study the divergence structure of the $\psi_{1,2}$ correlation functions, we only need to calculate a single loop (unlike for the case of the elementary $\sigma$ correlation function which is a geometric sum of two loop graphs). We will assume that an external momentum $p$ flows  into the right-most vertices in Fig.~\ref{fig:fig4} and to simplify the discussion we will set the energy and momentum of  the ``new"  incoming and outgoing particles (beyond the basic $U+2V \rightarrow 3 V$ reaction)  to zero. We will then only have to consider the internal one loop graphs at some arbitrary momentum $p$.  The expression for the one loop integral $I_1(p,t)$ is given by,
\ba
I_1(p,t) &=& 2! \int \prod_{i=1}^{2} \left(\frac{d^{d} p_i}{(2 \pi)^{d}}\right) e^{-[D_v(p_1^2+p_2^2) + 2 \mu]t} \nonumber \\
&\times& (2 \pi)^d \delta(p-p_1+p_2).
\ea
This can be calculated exactly and after Laplace transforming to $s$--space one gets,
\bq
I_1(p,s) = B_1  D_1 ^{-d/d_c} \frac{\Gamma(1+\epsilon/d_c)}{\epsilon/d_c} \left(s + \alpha_1 +  \frac{{D_1^2}}{4} p^2\right)^{-\epsilon/d_c},
\label{I1}
\eq
where now
\bq 
B_1 =  \frac{2!}{(4 \pi)^{d/d_c}}, \quad D_1 = 2 D_v , \quad \alpha_1 =   2 \mu.
\eq
Note that  for the single loop the critical dimension is now $d_c = 2$, and therefore $\epsilon = 2-d$. 
At $d=1$,  $\epsilon=1$  and we get the {\em finite} result:
\bq 
I_1(p,t) = B_1  D_1 ^{-1/2} \left(s + \alpha_1 +  \frac{{D_1^2}}{4} p^2\right)^{-1/2},
\eq
so that this process vanishes as $p^{-1}$ for large momenta. 

To evaluate 
 $I_1(p,s)$ in the critical dimension  $ d_c=2 $, we exponentiate the term in $s$ above and expand the exponential thus,
\ba
\left(s + \alpha_1 +  \frac{{D_1^2}}{4} p^2\right)^{-\epsilon/2} &=& e^{-(\epsilon/2) \ln(s')} \nonumber \\
&=& 1-\frac{\epsilon}{2} \ln s' + \Ord(\epsilon^2). 
\ea
Inserting the result of this expansion back into~\eqref{I1}, we immediately see that the only term that diverges as $\epsilon \rightarrow 0$ is the first term in the expansion.  In order to regulate this divergence we only need to renormalize the decay rate or so-called ``Mass" renormalization.  

We start by assuming that the one-loop integral $I_1$ corresponds to the correlation function for the composite field $\psi_1$.  Next identifying the zero energy-momentum piece of the dimensionless  propagator as $\alpha_1/M_1$ we get
\bq
I_1 (0) \equiv \frac{\alpha_1}{M_1} = {C_1} \left( \frac{1}{\epsilon} - \frac{1}{2} \ln \alpha_1  \right).
\eq
Consequently the renormalized version of the correlation function is now
\bq
\tilde{I_1}(p,s) =\frac{\alpha_1}{M_1}  - \frac{C_1}{2}   \ln {\frac{s'}{\alpha_1}}
\eq
where $C_1=  2  B_1  D_1 ^{-1} $. 

Thus in two dimensions, all  $n \to n$ processes are rendered finite by the introduction of only two new parameters $M_1, M_2$  corresponding to the decay rates for the two composite fields $\psi_1$ and $\psi_2$.  In terms of these two parameters  (and the renormalized coupling $g_R$) all the induced couplings can be determined. 

Finally, the full  expression for the Laplace transform of the $4 \rightarrow 4$ vertex function  of Fig. 5a  is just $g_R(\tilde s)^2 {I_1}(p,s)$,  where  at $d_c=2$ we need to replace $I_1$  by the regulated $\tilde{I_1}$.  It is not hard to see that for processes like Fig. 5b and extensions to the form $n U + 2 V \rightarrow (n-1) U + 3 V $
 this generalizes to 
\bq
g_R(\tilde s)^{n}\tilde{I_1}(p,s)^{n-1}.
\eq
%Interestingly in three dimensions the correlators are  again finite and do not need regulation. Explicitly the expression is
% \bq
%I_1(p,s) = -2B_1  D_1 ^{-3/2} \Gamma(1/2)
% \left(s + \alpha_1 +  \frac{{D_1^2}}{4} p^2\right)^{1/2}.
%\label{3dI}
%\eq

\section{Conclusions}

In this paper we presented a detailed analysis of the effects of intrinsic noise on a spatially extended reaction-diffusion chemical model~\eqref{reactions}. We found that remarkably, the short distance behavior of the system could be determined analytically by studying the model in the $U(1)$ symmetric phase of the field  theory~\cite{CGPM_2013}  corresponding to the chemical reactions described by Eq.~\eqref{reactions}. 
The symmetric phase of the field  theory reflects particle number conservation and consequently the only allowed graphs in its Feynman diagram representation are for $n \rightarrow n$, with $n \geq 3$.  The fluctuations due to the intrinsic noise leads to two types of  potentially divergent graphs in the theory. The first divergent graph, is a 2-loop graph which  first diverges in the critical dimension $d_c=1$.  This graph  we relate to the renormalization of the $3 \rightarrow 3$ reaction rate.  The second class of   divergent graphs first appear in $d=2$ in the induced $n \rightarrow n$ processes.   These we regulate using the standard technique of dimensional regularization.  We then investigated what (if any) are the new low energy (large spatio-temporal) parameters that need to be specified to define the correct finite and renormalized theory which includes the effects of the fluctuations. 

We find that one parameter is the critical dimension for the behavior of the reaction rate parameter $\lambda_0$, which happens to be $d_c=1$.  This reaction rate gets renormalized through a sequence of elastic collisions (two-loop graphs) that occur between the chemically relevant inelastic collision. As a result, it acquires a momentum dependence, which in the critical dimension we can specify by determining the reaction rate at late times. Equivalently, this is also determined by representing the elastic collisions as a composite three body state $\sigma = \phi_u \phi_v ^2$, and then determining its rate of decay. We also find that in $d_c = 1$ it is possible to distinguish between the situations where $\sigma$ is a bonafide composite state and the case where instead, it is an \emph{elementary} chemical with its own kinetic energy and decay terms. This is done by investigating the large momentum (short distance) behavior of the momentum dependent reaction rate. The decay rate in the version with an elementary $\sigma$ goes to zero in the high momentum UV limit faster than in the case where $\sigma$ is  composite. The point where the wave function renormalization of the elementary $ \sigma$  goes to zero, is where the two versions of the model are identical. 

Starting in two dimensions, two new parameters are needed to describe the system.  These can be thought of as being the decay rates (\emph{masses})  for the composite systems $\phi_{v}^2$ and $\phi_{u}\phi_{v}$.   To obtain these parameters one would need  to experimentally measure the reaction rates for two inelastic reactions such as $2U+2V \rightarrow U+ 3V$  and $3U + 2V \rightarrow 2U + 3V$.  Once that is done, all the induced $n \rightarrow n$  reaction rates can be calculated.  The renormalized  equation for the $\sigma$ field in two dimensions includes an induced \emph{free field} kinetic term.  In 3 dimensions the fluctuations have a further effect of changing the renormalized equation for the $\sigma$ field to one having higher spatio-temporal derivatives.    Thus in terms of the running coupling constant $g_R (\tilde s)$ as well as two measurable induced coupling constants, we have determined the effective field theory which results from the \emph{intrinsic} noise inherent in this chemical reaction diffusion model. 

We did not discuss the infrared properties of the theory with broken symmetry, which is the sector that relates directly to the chemistry.  To do so, one would follow a two step approach. First, one needs to determine the classical densities as a function of time and the classical response function which depends on both momentum and time.  Then we would make use of the running coupling constant found in this paper to determine the asymptotic behavior of the momentum dependent densities including fluctuations by solving the Callan-Symanzik equation~\cite{Callan_1970, Symanzik_1970}.  This approach is worked out in detail for the example of the process $3A  \rightarrow A$ in~\cite{Lee_1994} and extends the arguments standard in relativistic quantum field theory to this class of non-equilibrium models.
 
 \appendix
\section{Chemical master equation and many body formalism}
\label{sec:master}
In order to develop the master equation formalism for our system of chemical reactions~\eqref{reactions}, we first divide the space in which the reactions take place into a $d-$dimensional hyper-cubic lattice of cells and assume that we can treat each cell as a coherent entity. We assume the interactions occur locally at a single cell site and that there is also diffusion modeled as hopping between nearest neighbors. Assuming that the underlying processes are Markovian, they can be described by a probability distribution function $P(\mathbf{n_v, n_u} ,t)$ which gives the  probability to find the particle configuration $\mathbf{(n_v, n_u)}$ at time $t$.
Here  $\mathbf {n_i(t)} =( \{n_i(t) \}) $ is  a vector composition variable where $n_i$  represents the number of molecules of a species at site $i$. Following standard methods,   one obtains for the master equation for the chemical reactions in~\eqref{reactions} including diffusion 
\begin{widetext}
\ba \label{master2}
\frac{d}{dt}P(\mathbf{n_v, n_u},t) &=& \frac{D_v}{l^2}  \sum_{\langle i,j \rangle} \left[ (n_{v,j}+1)P(\ldots,n_{v,i}-1, n_{v,j}+1, \ldots, t) - n_{v,i}P \right] \nonumber \\
&+& \frac{D_u}{l^2}  \sum_{\langle i,j \rangle} \left[ (n_{u,j}+1)P(\ldots,n_{u,i}-1, n_{u,j}+1, \ldots, t) - n_{u,i}P \right]\nonumber \\
&+& \frac{\lambda}{2} \sum_{i} \left[(n_{v,i}-1)(n_{v,i}-2)(n_{u,i} +1)P(\ldots,n_{v,i}-1,\ldots, n_{u,i}+1, \ldots,t) -n_{v,i}(n_{v,i}-1)P \right] \nonumber \\
&+& \mu \sum_i  \left[ (n_{v,i}+1)P(\ldots, n_{v,i}+1, \ldots, t) - n_{v,i}P \right] + \nu \sum_i  \left[ (n_{u,i}+1)P(\ldots, n_{u,i}+1, \ldots, t) - n_{u,i}P \right] \nonumber \\
&+& f \sum_i  \left[ P(\ldots, n_{v,i}+1, \ldots, t) - P \right],
\ea
\end{widetext}
where $l$ is the characteristic length of the cell and $\langle \ldots \rangle$ denotes the sum over nearest neighbors.

The master equation~\eqref{master2} along with the sextic interaction shown in Fig.~\ref{fig:fig1} lends itself  to a many body description \cite{Doi76}, accomplished by the  introduction of an occupation number algebra with annihilation/creation operators
$\hat a_i, \hat a_i^\dag$ for $v$ and $\hat b_i, \hat b_i^\dag$ for $u$ at each site $i$. These operators obey the Bosonic commutation relations
\ba
\left[\hat a_i, \hat a^\dag _j \right] &=& \delta_{ij}, \quad \left[\hat b_i, \hat b^\dag _j \right]  = \delta_{ij}, \nonumber \\
\left[\hat a_i,\hat a _j \right] &=& 0, \quad~\left[\hat a_i^\dag,\hat a^\dag _j \right] =0,
\label{commute}
\ea
and define the occupation number operators $\hat n_{i,v} = \hat a_i ^{\dag} \hat a_i$ and $\hat n_{i,u} = \hat b_i ^{\dag} b_i$ satisfying the following eigenvalue equations:
\begin{equation}
\hat n_{i, v} | n _{i, v}\rangle = n_{i, v} | n _{i, v}\rangle,  \quad \hat n_{i, u} | n _{i, u}\rangle = n_{i, u} | n _{i, u}\rangle.  
\label{eigen}
\end{equation}
We next construct the state vector
\ba  \label{wavefunction}
|\Psi(t) \rangle =\sum_{\mathbf{n_v, n_u}}&& P(\mathbf{n_v, n_u} ,t) \nonumber \\
&& \times  \prod_i (\hat a^\dag_i)^{n_v ^i} (\hat b^\dag_i)^{n_u^i} |0 \rangle,
\ea
which upon differentiating with respect to time $t$, can be written in the suggestive form
\bq  \label{Schrodeq}
-\frac{\partial |\Psi(t) \rangle }{ \partial t} = H [\mathbf {\hat a ^{\dag}},\mathbf{\hat a},\mathbf{\hat b ^{\dag}},\mathbf{b}]|\psi_1(t) \rangle,
\eq
resembling the Schr\"odinger equation. Finally, taking  the time derivative of Eq. (\ref{wavefunction}) and comparing terms with the Hamiltonian in~\eqref{Schrodeq} we make the identification
\ba  \label{Hsym}
H &=& \frac{D _v}{l^2}  \sum_{\langle i,j \rangle}  (\hat a_i^\dag - \hat a_j^\dag)( \hat a_i- \hat a_j) + \mu \sum_i ( \hat a_i^\dag-1) \hat a_i   \nonumber \\
&&  + \frac{D _u}{l^2}  \sum_{\langle i,j \rangle}   ( \hat b_i^\dag-  \hat b_i^\dag) 
( \hat b_i- \hat b_j)  + \nu \sum_i( \hat b_i^\dag-1) \hat b_i \nonumber \\
&& -\frac{\lambda}{2}  \sum_i  \bigl[ \hat a_i^{\dag3} - \hat a_i^{\dag2} \hat b^{\dag} _i \bigr] \hat a_i^2  \hat b_i \nonumber \\
&& - f \sum_i(\hat b^\dag_i-1).
\ea
Having defined the space, the appropriate wave function and the  Hamiltonian, we next seek to evaluate the operator $ e^{- \tilde H t} $ using the path integral formulation. 
Following the standard procedure for obtaining the coherent state path integral~\cite{NO98, THVL_2005, Tauber_2007} to the GS system, letting the coherent state $\phi_v$  (related to the operator $a$)  represent $v$ and $\phi_u$ (related to the operator $b$)  represent $u$ we obtain
\bq
e^{-\tilde Ht} = \int \calD \phi_v \calD\phi_v^\star \calD \phi_u \calD\phi_u^\star e^{- S[\phi_v,\phi_v^\star,\phi_u,\phi_u^\star]}, \label{pathint}
\eq
where the action $S$  is given by
\ba
S = && \int dx \int_0^\tau dt \bigl[ \phi_v^\star \partial_t \phi_v + D_v \nabla \phi_v^\star \nabla \phi_v  +\phi_u^\star \partial_t \phi_u \nonumber \\
&&+ D_u \nabla \phi_u^\star \nabla \phi_u+ \mu (\phi_v^\star-1)  \phi_v + \nu (\phi_u^\star -1) \phi_u \nonumber \\
&& - f (\phi_u^\star -1)  
 - \frac {\lambda}{2}(\phi_v^\star-\phi_u^\star)\phi_v^{\star 2}  \phi_v^2 \phi_u  \bigr] .
 \label{action}
\ea
This is Eq. \eqref{actionunshifted} in the body of the text.

\end{document}